\documentclass[sigconf]{acmart}
\usepackage{epsfig}
\usepackage{graphicx}
\usepackage{algorithm}
\usepackage{rotating}

\usepackage{tabularx}
\usepackage{subfigure}
\usepackage{booktabs}
\usepackage{bm}
\usepackage{multirow}
\usepackage{enumitem}
\usepackage{balance}

\AtBeginDocument{%
  \providecommand\BibTeX{{%
    \normalfont B\kern-0.5em{\scshape i\kern-0.25em b}\kern-0.8em\TeX}}}

\copyrightyear{2022}
\acmYear{2022}
\setcopyright{acmcopyright}
\acmConference[WSDM '22] {Proceedings of the Fifteenth ACM International Conference on Web Search and Data Mining}{February 21--25, 2022}{Tempe, AZ, USA.}
\acmBooktitle{Proceedings of the Fifteenth ACM International Conference on Web Search and Data Mining (WSDM '22), February 21--25, 2022, Tempe, AZ, USA}
\acmPrice{15.00}
\acmISBN{978-1-4503-9132-0/22/02}
\acmDOI{10.1145/3488560.3498392}

\begin{document}
\fancyhead{}

\title{Personalized Transfer of User Preferences for Cross-domain Recommendation}


\author{Yongchun Zhu$^{1,3,4}$, Zhenwei Tang$^{1}$, Yudan Liu$^{4}$, Fuzhen Zhuang$^{2,5,*}$, Ruobing Xie$^{4}$, Xu Zhang$^4$, Leyu Lin$^4$ and Qing He$^{1,3}$}
\affiliation{%
 \institution{$^1$Key Lab of Intelligent Information Processing of Chinese Academy of Sciences (CAS), Institute of Computing Technology, CAS, Beijing 100190, China\\
 $^2$Institute of Artificial Intelligence, Beihang University, Beijing 100191, China\\
 $^3$University of Chinese Academy of Sciences, Beijing 100049, China\\
 $^4$WeChat Search Application Department, Tencent, China\\
 $^5$SKLSDE, School of Computer Science, Beihang University, Beijing 100191, China\\
 \{zhuyongchun18s, heqing\}@ict.ac.cn, lilvjosephtang@gmail.com, \{danydliu, ruobingxie, xuonezhang, goshawklin\}@tencent.com,zhuangfuzhen@buaa.edu.cn}\country{}}
\thanks{*Fuzhen Zhuang is the corresponding author.}

\renewcommand{\shortauthors}{Trovato and Tobin, et al.}

\begin{abstract}
  Cold-start problem is still a very challenging problem in recommender systems. Fortunately, the interactions of the cold-start users in the auxiliary source domain can help cold-start recommendations in the target domain. How to transfer user's preferences from the source domain to the target domain, is the key issue in Cross-domain Recommendation (CDR) which is a promising solution to deal with the cold-start problem. Most existing methods model a common preference bridge to transfer preferences for all users.
  Intuitively, since preferences vary from user to user, the preference bridges of different users should be different.
  Along this line, we propose a novel framework named Personalized Transfer of User Preferences for Cross-domain Recommendation (PTUPCDR). Specifically, a meta network fed with users' characteristic embeddings is learned to generate personalized bridge functions to achieve personalized transfer of preferences for each user. To learn the meta network stably, we employ a task-oriented optimization procedure. With the meta-generated personalized bridge function, the user's preference embedding in the source domain can be transformed into the target domain, and the transformed user preference embedding can be utilized as the initial embedding for the cold-start user in the target domain.  Using large real-world datasets, we conduct extensive experiments to evaluate the effectiveness of PTUPCDR on both cold-start and warm-start stages. The code has been available at \url{https://github.com/easezyc/WSDM2022-PTUPCDR}.
\end{abstract}




\maketitle

\section{Introduction}

Recommender systems are playing more and more important roles in web and mobile applications. In recent years, recommender systems have attracted a vast amount of interest from industries and academia, and researchers have conducted a great deal of research to improve the recommendation performance~\cite{koren2009matrix,covington2016deep}. However, most of these recommender systems are hard to provide satisfying recommendations for new users, i.e., cold-start users, who have no historical interactions.

\begin{figure}[t!]
\centering
\begin{minipage}[b]{1\linewidth}
\centering
\includegraphics[width=\linewidth]{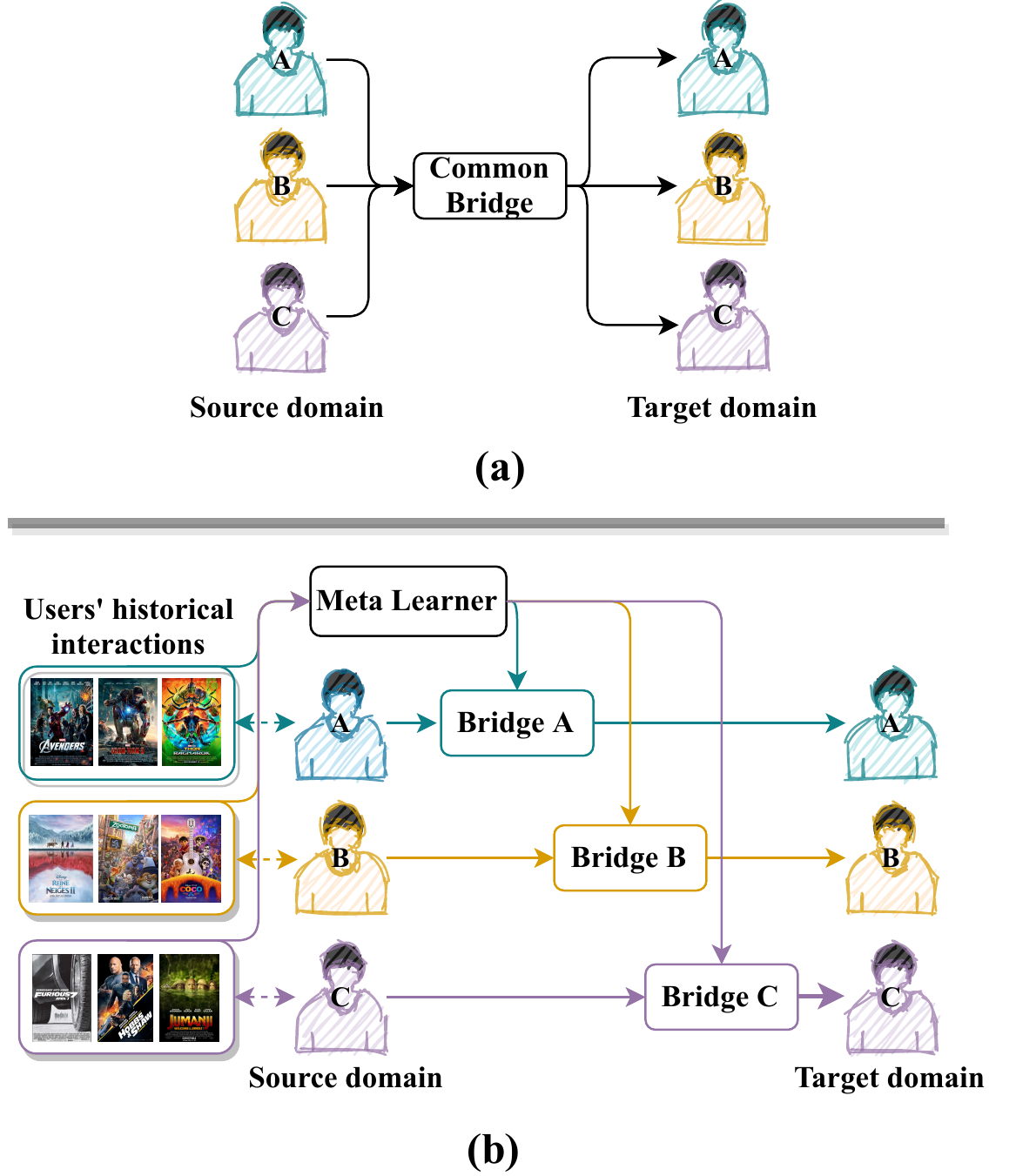}\vspace{-0.2cm}
\end{minipage}
\caption{(a) In existing CDR methods: all users share the common bridge function. (b) The proposed PTUPCDR utilizes a meta network to generate personalized bridge functions for each user.}\label{fig:compare}\vspace{-0.3cm}
\end{figure} 

Cross-domain Recommendation (CDR)~\cite{singh2008relational} which aims to transfer knowledge from an informative source domain to the target domain is a promising solution to alleviate the cold-start problem. The core task of CDR is to bridge user's preferences in the source domain and the target domain, also called preference transfer~\cite{cheng2020catn}. To achieve preference transfer, many existing CDR methods~\cite{man2017cross,zhu2018deep,fu2019deeply,kang2019semi} assume that all users share the same relationships between user preferences in the source domain and the target domain, and learn a common preference bridge shared by all users, as shown in Figure~\ref{fig:compare} (a).

In practice, due to individual differences, the complex relationships between the user preferences of the source and target domains vary from user to user. Hence, it is hard for a single bridge to capture such complicated and various relationships, which may degrade these CDR methods' performance. To alleviate the drawback, it is necessary to use personalized bridges to model various relationships between user preferences in different domains. In other words, the process of preference transfer should be personalized.

Along this line, we propose a novel framework named Personalized Transfer of User Preferences for Cross-domain Recommendation (PTUPCDR). PTUPCDR learns a meta network that takes users' characteristic embeddings in the source domain as input and generates personalized bridges for each user, as shown in Figure~\ref{fig:compare}(b). 
The generated bridge functions can be viewed as a model parameterized by the learned meta network. Note that the personalized bridge functions which depend on the users' characteristics vary from user to user, so the process of the preference transfer is personalized, which can capture preference relationships between different domains better than existing methods. 
After training, we feed user embeddings in the source domain into the meta-generated personalized bridge functions and obtain the transformed user embeddings. The transformed user embeddings are utilized as the initial embeddings in the target domain. With the initial embeddings, our method is effective for cold-start users who have no interactions in the target domain. 

In practice, a high-level meta network is hard to optimize~\cite{munkhdalai2017meta,shen2018neural}, and how to optimize the meta network is another challenge. To learn the bridge function, existing bridge-based methods~\cite{man2017cross,zhu2018deep,fu2019deeply,kang2019semi} adopt a mapping-oriented optimization procedure to directly minimize the distance between the transformed users' embeddings from the informative source domain and the user's embedding in the target domain. In other words, with such an optimization procedure, the bridge function is sensitive to the quality of the users' embeddings. In practical recommender systems, it is pretty hard to learn reasonable embeddings~\cite{pan2019warm,zhu2021learning} for all users, which limits the performance of the bridge function learned with the mapping-oriented optimization. In addition, we find it is hard to learn the meta network with the mapping-oriented optimization. Thus, to train the meta network, we take a task-oriented optimization procedure, which skips the users' embeddings in the target domain and directly utilizes the rating task as the optimization goal.

Most existing works in the literature~\cite{man2017cross,zhu2018deep,fu2019deeply,kang2019semi,cheng2020catn} only testify their effectiveness by applying their methods upon simple base models (Matrix Factorization) in the extreme cold-start stage (users have no interactions in the target domain). Since such settings are far from real-world scenarios, we further explore how to use PTUPCDR in more practical scenarios to validate the compatibility and utility of PTUPCDR in real-world recommendations, e.g., warm-start scenarios, more complicated base models. Experimental results demonstrate that our proposed PTUPCDR is of good compatibility and utility in real-world recommendations.

The main contributions of our work are summarized into three folds:
\begin{itemize}
\item To solve the cold-start problem in CDR, we propose a novel method named PTUPCDR, utilizing a meta network to generate personalized bridge functions for each user, given the encoded users' characteristics in the source domain.
\item To learn the meta network stably, we employ a task-oriented optimization procedure to alleviate the side effects of unreasonable users' embeddings.
\item We conduct extensive experiments on three cross-domain tasks using Amazon review dataset to demonstrate the effectiveness and robustness of PTUPCDR for not only cold-start scenarios but also warm-start scenarios, while existing methods only testify their effectiveness in the cold-start scenarios.
\end{itemize}

\section{Related Work}

\subsection{Cross-domain Recommendation} 
Transfer learning aims to leverage knowledge from a source domain to improve the learning performance or minimize the number of labeled examples required in a target domain~\cite{zhuang2020comprehensive,wang2021generalizing}, which led to an interest in many
area, e.g., computer vision~\cite{wang2018visual,zhu2019aligning,zhu2020deep}, natural language processing~\cite{xi2020domain,nan2021mdfend}. Inspired by transfer learning, CDR is a promising solution to alleviate data sparsity and the cold-start problem in the target domain with the help of the auxiliary (source) domain. At the very beginning, CMF~\cite{singh2008relational} assumes a shared global user embedding matrix for all domains, and it factorizes matrices from multiple domains simultaneously. CST~\cite{pan2010transfer} utilizes the user embedding in the source domain to initialize the embedding in the target domain and restricts them from being closed.

In recent years, researchers proposed many deep learning-based models to enhance knowledge transfer~\cite{hu2018conet,he2018robust,xie2021contrastive,zhu2021learning,hao2021adversarial,xi2021modeling}. CoNet~\cite{hu2018conet} transfers and combines the knowledge by using cross-connections between feed-forward neural networks. MINDTL~\cite{he2018robust} combines the CF information of the target-domain with the rating patterns extracted from a cluster-level rating matrix in the source-domain. DDTCDR~\cite{li2020ddtcdr} develops a novel latent orthogonal mapping to extract user preferences over multiple domains while preserving relations between users across different latent spaces. 

Another group of CDR methods focus on bridging user preferences in different domains~\cite{pan2010transfer,man2017cross,zhu2018deep,kang2019semi,zhang2020learning,cheng2020catn,zhu2021transfer}, which is the most related work. CST~\cite{pan2010transfer} utilizes the user embedding learned in the source domain to initialize the user embedding in the target domain and restricts them to being closed. Some methods~\cite{man2017cross,kang2019semi,cheng2020catn,zhu2021transfer} explicitly model the preference bridge. Our  study falls into the this bridge-based category. However, to the best of our knowledge, all of the bridge-based CDR works in the literature learn a shared bridge function for all users, while our PTUPCDR is the first to learn personalized bridges for each user. 

\subsection{Cold-start Recommendation}
Providing recommendations for new users or items is challenging in recommender systems, also named the cold-start problem. There are two kinds of methods to solve the cold-start problems. The first type actively solves cold-start by designing a decision making strategy, such as using contextual-bandits~\cite{li2010contextual,pan2019policy}.

This paper belongs to the second type, which utilizes auxiliary information to help the cold-start stage. There are various kinds of auxiliary information could be exploited to improve cold-start recommendation performance, e.g., user attributes~\cite{seroussi2011personalised,li2019zero}, item attributes~\cite{mo2015image,xie2020internal,zhu2021learning}, knowledge graphs~\cite{wang2018ripplenet}, samples in an auxiliary domain~\cite{man2017cross}, etc.  Usually, with samples in an auxiliary domain, the CDR methods can achieve much better results than other cold-start methods. Thus, in this paper, following most CDR work~\cite{man2017cross,kang2019semi}, we only compare our method with CDR approaches.

\subsection{Meta Learning}
It is also named learning to learn, aiming to improve novel tasks' performance by training on similar tasks. There are various meta learning methods, e.g., metric-based methods~\cite{vinyals2016matching,snell2017prototypical}, gradient-based methods~\cite{finn2017model}, and parameter-generating based methods~\cite{munkhdalai2017meta}. The proposed PTUPCDR falls into the third group, which utilizes a meta learner to predict networks' parameters. Recently, researchers proposed many meta-based methods~\cite{pan2019warm,lee2019melu,zhu2021learning,zhu2021learningkdd} to improve recommender systems' performance. However, most of them fall into gradient-based methods and focus on single-domain recommendations, while we focus on cross-domain recommendations. The most related work is TMCDR~\cite{zhu2021transfer} which utilizes meta-learning in CDR. However, TMCDR also trains a common bridge as existing bridge-based methods do.

\section{Model}
\subsection{Problem Setting}
In CDR, we have a source domain and a target domain. Each domain has a user set $\mathcal{U} = \{u_1, u_2, ...\}$, an item set $\mathcal{V} = \{v_1, v_2, ... \}$, and a rating matrix $\mathcal{R}$. $r_{ij} \in \mathcal{R}$ denotes the interaction between user $u_i$ and item $v_j$. To distinguish these two domains, we denote the user, item sets, and the rating matrix of the source domain as $\mathcal{U}^s, \mathcal{V}^s, \mathcal{R}^s$, while $\mathcal{U}^t, \mathcal{V}^t, \mathcal{R}^t$ for the target domain. We define the overlapping users between the two domains as $\mathcal{U}^o = \mathcal{U}^s \cap \mathcal{U}^t$. In contrast, $\mathcal{V}^s$ and $\mathcal{V}^t$ are disjoint, which means there is no shared item between the two domains. 

\begin{figure*}[t!]
    \centering
\begin{minipage}[b]{0.9\linewidth}
\centering
\includegraphics[width=1\linewidth]{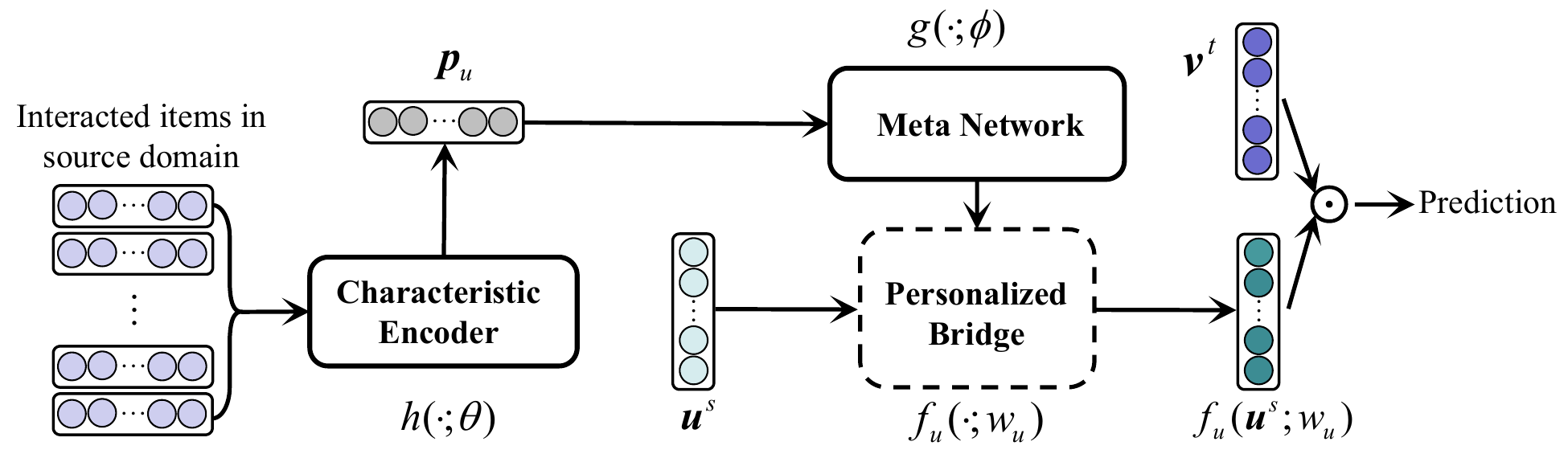}
\end{minipage}
\caption{Personalized Transfer of User Preferences for Cross-domain Recommendation (PTUPCDR) utilizes a meta network with users' characteristic embeddings in the source domain as input to generate personalized bridge functions for each user. Then, with the personalized bridge function, we can obtain the transformed user's embeddings as the initial embeddings.}\label{fig:network}
\end{figure*}

In latent factor models, the users and items are transformed into dense vectors, also called factors or embeddings. In this paper, $\bm{u}^d_i \in \mathbb{R}^{k}$ and $\bm{v}^d_j  \in \mathbb{R}^{k}$ denote the embeddings of the user $u^d_i$ and item $v^d_j$, respectively, where $k$ denotes the dimensionality of 
embeddings and $d \in \{s ,t\}$ represents the domain label. For each user $u_i$, we denote the list of her sequential interaction items in source domain by $\mathcal{S}_{u_i} = \{v^s_{t_1}, v^s_{t_2}, \cdots, v^s_{t_n}\}$, where $n$ denotes the number of interacted items and $v^s_{t_n}$ denotes the interacted item in the source domain at timestamp $t_n$.

\subsection{Characteristic Encoder}\label{sec:attention}
The first step to generate the personalized bridge function is to capture users' personalized transferable characteristics from interacted items. However, cold-start users have no interacted item in the target domain. Thus, it is essential to exploit the interacted items $\mathcal{S}$ in the source domain. Note that we need to find the transferable characteristics which are helpful for knowledge transfer.

Intuitively, various items have different contributions to knowledge transfer. The attention mechanism~\cite{2017Attentional,zhou2018deep} allows different parts to contribute differently when compressing them to a single representation. Therefore, we propose to employ the attention mechanism on item embeddings by performing a weighted sum:
\begin{equation}
    \bm{p}_{u_i} = \sum_{v^s_j \in \mathcal{S}_{u_i}} a_j \bm{v}^s_j,
\end{equation}
where $\bm{p}_{u_i} \in \mathbb{R}^{k}$ denotes the transferable characteristic embedding of user $u_i$, and $a_j$ is the attention score for item $v_j$, which can be interpreted as the importance of $v_j$ in predicting the personalized bridge function. For the target domain, an irrelevant item would has little help for personalized bridge functions of all users. Thus, we learn the attention score from the items' embeddings by an attention network. Formally, the attention network is defined as:
\begin{equation}
    \begin{split}
        a'_j &= h(\bm{v}_j;\theta),\\
        a_j &= \frac{\exp(a'_j)}{\sum_{v^s_l \in \mathcal{S}_{u_i}} \exp(a'_l)},
    \end{split}
\end{equation}
where $h(\cdot)$ denotes the attention network, and $\theta$ denotes the parameters of $h(\cdot)$. In this paper, $h(\cdot)$ is a two-layer feed-forward network. Note that the normalized attention score $a_j$ is beneficial to find the useful interacted items for a specific user. After that, we could utilize each user's characteristics as input to guide the generation of personalized bridge function.

\setlength{\textfloatsep}{7pt}
\begin{algorithm} [t] 
    \caption{Personalized Transfer of User Preferences for CDR (PTUPCDR)}\label{alg}
    \flushleft{\hspace*{0.02in}\textbf{Input}: $\mathcal{U}^s, \mathcal{U}^t, \mathcal{V}^s, \mathcal{V}^t$, $\mathcal{U}^o$, $\mathcal{R}^s, \mathcal{R}^t$
    
    \hspace*{0.02in}\textbf{Input}: Meta network $g_\phi$.
    
    \hspace*{0.02in}\textbf{Input}: Characteristic encoder $h_\theta$.
    
    \hspace*{0.02in}\textbf{Pre-training Stage}:
    
    \hspace*{0.06in}1. Learning a source model which contains $\bm{u}^s, \bm{v}^s$.
        
    \hspace*{0.06in}2. Learning a target model which contains $\bm{u}^t, \bm{v}^t$.
    
    \hspace*{0.02in}\textbf{Meta Stage}: 
    
    \hspace*{0.06in}3. Learning a characteristic encoder $h_\theta$ and a meta network $g_\phi$ by minimizing Equation~(\ref{eq:taskoriented}).
    
    \hspace*{0.02in}\textbf{Initialization Stage}: 
    
    \hspace*{0.06in}4. For a cold-start user $u^t$ in the target domain, we use the transformed embedding $f_{u_i}(\bm{u}^s_i;\bm{w}_{u_i})$ as the user's initialized embedding in the target domain.}
    
\end{algorithm}

\subsection{Meta Network}\label{sec:meta}
We have mentioned that the users' relationships between preferences of different domains vary from user to user. In other words, the process of preference transfer needs to be personalized. Intuitively, there exists a certain connection between the preference relationship and the user's characteristics. Based on this intuition, we propose a meta network which takes the user's transferable characteristics as input, and then generates a personalized bridge function between the user's embeddings in the source and target domains. The proposed meta network is formulated as:
\begin{equation}
    \bm{w}_{u_i} = g(\bm{p}_{u_i}; \phi),
\end{equation}
where $g(\cdot)$ is the meta network, which is parameterized by $\phi$. In this paper, the meta network is a two-layer feed-forward network. The $\bm{w}_{u_i}$ is a vector whose size depends on the structure of the bridge function. The personalized bridge function is formulated as:
\begin{equation}
    f_{u_i}(\cdot;\bm{w}_{u_i}),
\end{equation}
which utilizes $\bm{w}_{u_i}$ as the parameters of bridge function $f(\cdot)$. The bridge function can be defined as any structure. In this paper, for simplicity, we use a linear layer as $f(\cdot)$ following EMCDR~\cite{man2017cross,kang2019semi}. Thus, to fit the size of bridge's parameters, we reshape the vector $\bm{w}_{u_i} \in \mathbb{R}^{k^2}$ into a matrix $\bm{w}_{u_i} \in \mathbb{R}^{k \times k}$. Note that the $\bm{w}_{u_i}$ is used as the parameters of the bridge functions rather than input. The generated bridge function depends on user's characteristics and varies from user to user, and we call it the personalized bridge function. 

With the personalized bridge function, we can obtain the personalized transformed user's embeddings:
\begin{equation}
    \hat{\bm{u}}_i^t = f_{u_i}(\bm{u}^s_i;\bm{w}_{u_i}),
\end{equation}
where $\bm{u}^s_i$ denotes the embedding of user $u_i$ in the source domain, and $\hat{\bm{u}}_i^t$ represents the transformed embedding. Finally, we can utilize the transformed embedding $\hat{\bm{u}}_i^t$ for prediction.

\subsection{Task-oriented Optimization}\label{sec:taskoriented}

To train the meta network and characteristic encoder, we can minimize the distance using the \textbf{mapping-oriented optimization} procedure following existing bridge-based methods~\cite{man2017cross,zhu2018deep,fu2019deeply,kang2019semi}:
\begin{equation}
    \mathcal{L} = \sum_{u_i \in \mathcal{U}^o} || \hat{\bm{u}}^t_i - \bm{u}_i^t ||^2,\label{mappingoriented}
\end{equation}
where $ \hat{\bm{u}}^t_i$ denotes the transformed user embedding from $\bm{u}^s_i$ in the source domain, and $\bm{u}_i^t$ denotes the user embedding in target domain. The mapping-oriented optimization procedure would bring the transformed embedding $\hat{\bm{u}}_i^t$ close to the target embedding $\bm{u}_i^t$. 

However, since some users only have limited interactions, the user's embedding $\bm{u}_i^t$ may be not reasonable and accurate enough. Learning towards the relatively unreasonable embeddings would lead to negative impact on the model. Thus, we propose a task-oriented optimization to train the meta network and characteristic encoder. The task-oriented training procedure directly utilizes the performance of the ultimate recommendation task as the optimization goal. In this paper, we focus on rating task, so the task-oriented loss can be formulated as:
\begin{equation}
    \min_{\theta, \phi} \frac{1}{|\mathcal{R}^t_o|} \sum_{r_{ij} \in \mathcal{R}^t_o} (r_{ij} - f_{u_i}(\bm{u}^s_i;\bm{w}_{u_i})\bm{v}_j)^2,\label{eq:taskoriented}
\end{equation}
where $\mathcal{R}^t_o = \{r_{ij}| u_i \in \mathcal{U}^o, v_j \in \mathcal{V}^t\}$ denotes the interactions of overlapping users in the target domain. 

\begin{table*}[t]
  \centering
  \caption{Statistics of the cross-domain tasks ($Overlap$ denotes the number of overlapping users).}
    \setlength{\tabcolsep}{2.8mm}
    \begin{tabular}{c||cc||cc||ccc||cc}
    \toprule
    \multirow{2}[2]{*}{\textbf{CDR Tasks}} & \multicolumn{2}{c||}{\textbf{Domain}} & \multicolumn{2}{c||}{\textbf{Item}} & \multicolumn{3}{c||}{\textbf{User}} & \multicolumn{2}{c}{\textbf{Rating}} \\
          & Source & Target & Source & Target & Overlap & Source & Target & Source & Target \\
    \midrule
    \textbf{Task1} & Movie & Music & 50,052  & 64,443  & 18,031  & 123,960  & 75,258  & 1,697,533  & 1,097,592  \\
    \textbf{Task2} & Book  & Movie & 367,982  & 50,052  & 37,388  & 603,668  & 123,960  & 8,898,041  & 1,697,533  \\
    \textbf{Task3} & Book  & Music & 367,982  & 64,443  & 16,738  & 603,668  & 75,258  & 8,898,041  & 1,097,592  \\
    \bottomrule
    \end{tabular}
  \label{tab:count}%
\end{table*}%

Compared with the mapping-oriented procedure, task-oriented optimization has two advantages: (1) The task-oriented optimization can alleviate the effects of unreasonable embeddings. It directly uses the rating data, which is ground truth rather than approximate intermediate results. (2) The task-oriented learning procedure has more training samples, which can avoid overfitting. For example, with $N$ overlapping users, each user has $M$ ratings. The mapping-oriented process learns the mapping function with $|\mathcal{U}^o = N|$ samples as Equation~(\ref{mappingoriented}), while the task-oriented learning procedure utilizes the $|\mathcal{R}^t_o| = M \times N$ user-item rating as Equation~(\ref{eq:taskoriented}).

\subsection{Overall Procedure}
The overall structure of PTUPCDR is shown in Figure~\ref{fig:network}. The training procedure can be divided into three steps: pre-training, meta and initialization stages, as see Algorithm~\ref{alg}. After training, the method can work for both cold-start and warm-start stages.

\textbf{Pre-training stage}: This step is to learn latent spaces for each domain, respectively. The loss function is formulated as:
\begin{equation}
    \min_{\bm{u}, \bm{v}} \frac{1}{|\mathcal{R}|} \sum_{r_{ij}\in \mathcal{R}} (r_{ij} - \bm{u}_i \bm{v}_j)^2,
\end{equation}
where $|\mathcal{R}|$ denotes the number of ratings. After the pre-training step, we can obtain the pre-trained embeddings $\bm{u}^s, \bm{u}^t, \bm{v}^s, \bm{v}^t$. 

\textbf{Meta stage}: The existing methods directly train a common bridge function, while PTUPCDR trains the characteristic encoder and the meta network. The characteristic encoder and the meta network are optimized with Equation~(\ref{eq:taskoriented}).


\textbf{Initialization stage}: When a new user comes (CDR assumes the new user has some interactions in the source domain), we use the transformed embedding $\hat{\bm{u}}_i^t = f_{u_i}(\bm{u}^s_i;\bm{w}_{u_i})$ to initialize the new user's embedding in the target domain.

\textbf{Test stage}: For the extreme cold-start users who have no interactions in the target domain, directly utilize the initial embedding $\hat{\bm{u}}_i^t = f_{u_i}(\bm{u}^s_i;\bm{w}_{u_i})$ for prediction. For the warm-start users who have some interaction in the target domain, it is convenient to fine-tune the initial embeddings with new interactions, and utilize the fine-tuned embeddings for prediction.

\begin{table*}[t]
  \centering
  \caption{Cold-start results (MAE and RMSE) of 3 cross-domain tasks. We report the mean results over five runs. Best results are in boldface. $*$ indicates $0.05$ level, paired t-test of PTUPCDR vs. the best baselines. $Improve$ denotes relative improvement over the best baseline.}
    \setlength{\tabcolsep}{3mm}
    \begin{tabular}{c||c||c||cccccc||c}
    \toprule
          & \textit{\textbf{$\beta$}} & \textit{\textbf{Metric}} & \textbf{TGT} & \textbf{CMF} & \textbf{DCDCSR} & \textbf{SSCDR} & \textbf{EMCDR} & \textbf{PTUPCDR} & \textbf{Improve} \\
    \midrule
    \multirow{6}[6]{*}{\textbf{Task1}} & \multirow{2}[2]{*}{\textit{20\%}} & \textit{MAE} & 4.4803  & 1.5209  & 1.4918  & 1.3017  & 1.2350  & \textbf{1.1504*} & 6.86\% \\
          &       & \textit{RMSE} & 5.1580  & 2.0158  & 1.9210  & 1.6579  & 1.5515  & \textbf{1.5195 } & 2.06\% \\
\cmidrule{2-10}          & \multirow{2}[2]{*}{\textit{50\%}} & \textit{MAE} & 4.4989  & 1.6893  & 1.8144  & 1.3762  & 1.3277  & \textbf{1.2804*} & 3.57\% \\
          &       & \textit{RMSE} & 5.1736  & 2.2271  & 2.3439  & 1.7477  & 1.6644  & \textbf{1.6380 } & 1.59\% \\
\cmidrule{2-10}          & \multirow{2}[2]{*}{\textit{80\%}} & \textit{MAE} & 4.5020  & 2.4186  & 2.7194  & 1.5046  & 1.5008  & \textbf{1.4049*} & 6.39\% \\
          &       & \textit{RMSE} & 5.1891  & 3.0936  & 3.3065  & 1.9229  & 1.8771  & \textbf{1.8234*} & 2.86\% \\
    \midrule
    \multirow{6}[6]{*}{\textbf{Task2}} & \multirow{2}[2]{*}{\textit{20\%}} & \textit{MAE} & 4.1831  & 1.3632  & 1.3971  & 1.2390  & 1.1162  & \textbf{0.9970*} & 10.68\% \\
          &       & \textit{RMSE} & 4.7536  & 1.7918  & 1.7346  & 1.6526  & 1.4120  & \textbf{1.3317*} & 5.69\% \\
\cmidrule{2-10}          & \multirow{2}[2]{*}{\textit{50\%}} & \textit{MAE} & 4.2288  & 1.5813  & 1.6731  & 1.2137  & 1.1832  & \textbf{1.0894*} & 7.93\% \\
          &       & \textit{RMSE} & 4.7920  & 2.0886  & 2.0551  & 1.5602  & 1.4981  & \textbf{1.4395*} & 3.91\% \\
\cmidrule{2-10}          & \multirow{2}[2]{*}{\textit{80\%}} & \textit{MAE} & 4.2123  & 2.1577  & 2.3618  & 1.3172  & 1.3156  & \textbf{1.1999*} & 8.80\% \\
          &       & \textit{RMSE} & 4.8149  & 2.6777  & 2.7702  & 1.7024  & 1.6433  & \textbf{1.5916*} & 3.15\% \\
    \midrule
    \multirow{6}[6]{*}{\textbf{Task3}} & \multirow{2}[2]{*}{\textit{20\%}} & \textit{MAE} & 4.4873  & 1.8284  & 1.8411  & 1.5414  & 1.3524  & \textbf{1.2286*} & 9.15\% \\
          &       & \textit{RMSE} & 5.1672  & 2.3829  & 2.2955  & 1.9283  & 1.6737  & \textbf{1.6085*} & 3.90\% \\
\cmidrule{2-10}          & \multirow{2}[2]{*}{\textit{50\%}} & \textit{MAE} & 4.5073  & 2.1282  & 2.1736  & 1.4739  & 1.4723  & \textbf{1.3764*} & 6.51\% \\
          &       & \textit{RMSE} & 5.1727  & 2.7275  & 2.6771  & 1.8441  & 1.8000  & \textbf{1.7447*} & 3.07\% \\
\cmidrule{2-10}          & \multirow{2}[2]{*}{\textit{80\%}} & \textit{MAE} & 4.5204  & 3.0130  & 3.1405  & 1.6414  & 1.7191  & \textbf{1.5784*} & 3.84\% \\
          &       & \textit{RMSE} & 5.2308  & 3.6948  & 3.5842  & 2.1403  & 2.1119  & \textbf{2.0510*} & 2.88\% \\
    \bottomrule
    \end{tabular}%
  \label{tab:result}%
\end{table*}%

\section{Experiments}
We conduct experiments to answer the following research questions: \textbf{RQ1} Why we need an auxiliary domain and why we need to introduce CDR? How does PTUPCDR perform in extremely cold-start scenarios comparing to state-of-the-art models with a CDR perspective?
\textbf{RQ2} How does PTUPCDR perform in more practical scenarios of real-world recommendations?
\textbf{RQ3} Why could PTUPCDR perform better?

\subsection{Experimental Settings}
\textbf{Datasets.}
Following most existing methods~\cite{kang2019semi,cheng2020catn,zhu2021transfer}, A real-world public dataset is adopted for experiments, namely the Amazon review dataset~\footnote{http://jmcauley.ucsd.edu/data/amazon/}. Specifically, we use the Amazon-5cores dataset in which each user or item has at least five ratings. 

Following~\cite{kang2019semi,cheng2020catn}, we choose 3 popular categories out of 24 in total: movies\_and\_tv (Movie), cds\_and\_vinyl (Music), and books (Book). We define 3 CDR tasks as Task 1: Movie $\rightarrow$ Music, Task 2: Book $\rightarrow$ Movie, and Task 3: Book $\rightarrow$ Music. As the details listed in Table~\ref{tab:count}, the number of ratings of the source domain is significantly large than the one in the target domain. While many existing works only select a part of the dataset for evaluation, we directly use all data to simulate the real-world application. 

\textbf{Evaluation Metrics.}
Amazon review dataset contains rating data (0 - 5 score). Following~\cite{man2017cross,cheng2020catn} we adopt Mean Absolute Error (MAE) and Root Mean Square Error (RMSE) as the metrics.

\textbf{Baselines.}
Since PTUPCDR falls into the bridge-based methods for CDR, we mainly compare PTUPCDR with the bridge-based methods. Therefore, we choose the following methods as baselines for comparison. 1) TGT denotes the target MF model, which is trained only using target domain data. 2) CMF~\cite{singh2008relational} is an extension of MF. In CMF, the embeddings of users are shared across the source and target domains. 3) EMCDR~\cite{man2017cross} is a popular CDR method for cold-start. It adopts Matrix Factorization (MF) to learn embeddings first and then utilize a network to bridge the user embeddings from the auxiliary domain to the target domain. 4) DCDCSR~\cite{zhu2018deep} falls into the bridge-based methods, which considers the rating sparsity degrees of individual users in different domains. 5) SSCDR~\cite{kang2019semi} is a semi-supervised bridge-based method.

\textbf{Implementation Details.}\label{sec:implementation}
We implement our framework and the baselines using PyTorch. For each task and method, the initial learning rate for the Adam~\cite{kingma2014adam} optimizer are tuned by grid searches within \{0.001, 0.005, 0.01, 0.02, 0.1\}. In addition, we set the dimension of embeddings as 10. For all methods, we set mini-batch size of 512. We employ the same fully connected layer to facilitate comparison for the cross-domain bridge functions of EMCDR, DCDCSR, SSCDR, and PTUPCDR. Note that the personalized bridge function of PTUPCDR is generated by the meta network. The meta network is a two-layer network with hidden units $2 \times k$, where $k$ denotes the embedding dimension, and the output's dimension of the meta network is $k \times k$. The attention network is a two-layer network with $k$ hidden units.

\begin{figure*}[t!]
\centering
\begin{minipage}[b]{1\linewidth}
\centering
\includegraphics[width=.92\linewidth]{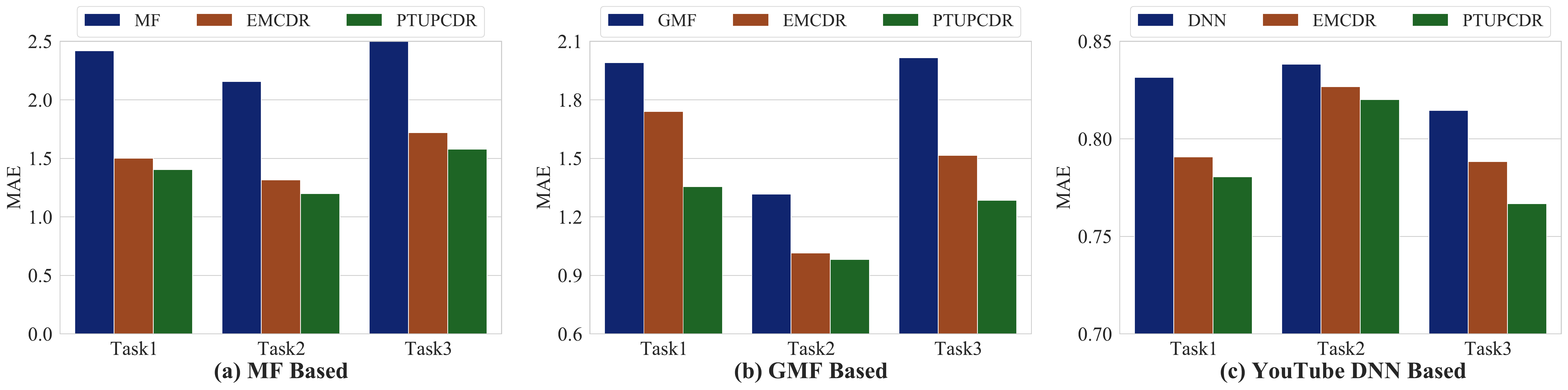}
\end{minipage}\vspace{-0.1cm}
\caption{Generalization experiments: applying EMCDR and PTUPCDR upon three base models (a) MF, (b) GMF, and (c) YouTube DNN, and show the averaged results over five runs. }\label{fig:generalization}
\end{figure*}

\begin{figure*}[t!]
\centering
\begin{minipage}[b]{1\linewidth}
\centering
\includegraphics[width=.92\linewidth]{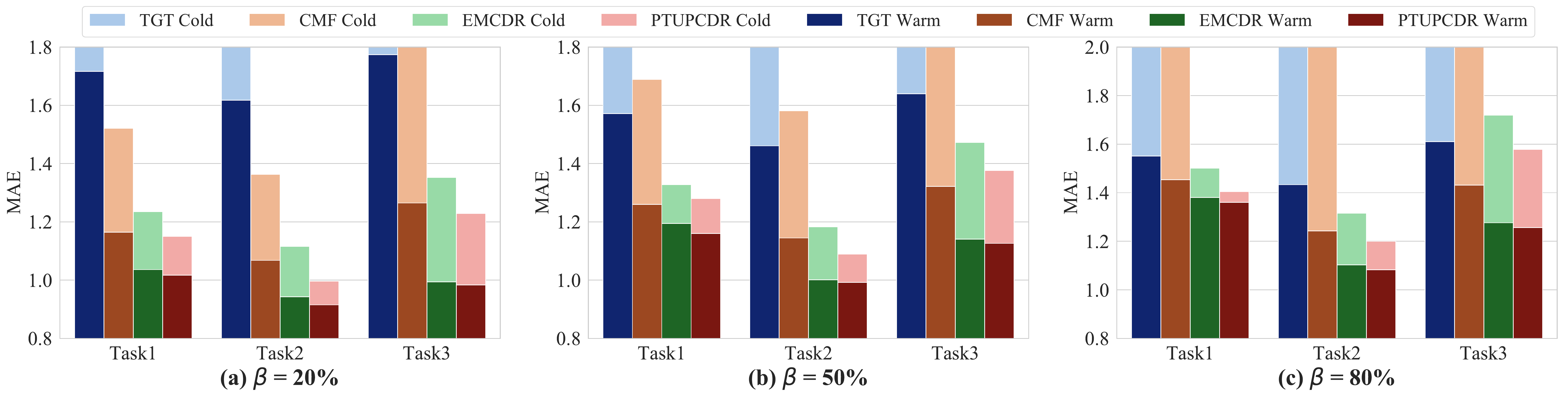}
\end{minipage}\vspace{-0.1cm}
\caption{Warm-start experiments on TGT, CMF, EMCDR, and PTUPCDR for different proportions of test (cold-start) users $\beta$: (a) $\beta = 20\%$, (b) $\beta = 50\%$, and (c) $\beta = 80\%$. The light-colored histograms represent the performance of extreme cold-start scenario, while the dark-colored histograms represent the warm-start scenario.}\label{fig:warm}
\end{figure*}

Following~\cite{man2017cross}, to evaluate the effectiveness of the proposed PTUPCDR on cross-domain recommendation, we randomly remove all the ratings of a fraction of overlapping users in the target domain and regard them as test users, and the samples of other overlapping users are used for training the bridge function. In our experiments, we set the proportions of test (cold-start) users $\beta$ as 80\%, 50\%, and 20\% of the total overlapping users, respectively. For the cold-start experiments in Table~\ref{tab:result} and Figure~\ref{fig:generalization}, all ratings of test users are used as the test set. For the warm-start experiments in Figure~\ref{fig:warm}, we divide the ratings of each test user into two parts with a ratio of 1:1. Note that we take the sequential timestamps into account to avoid information leakage. We use the first part as the cold-start set, and the other as the warm-start set. The process of model evaluation can be divided into three step: 1) Train the model on the training set. 2) Test extremely cold-start performance on the cold-start set. 3) Fine-tune the target model with the cold-start set, and evaluate the warm-start performance on the warm-start set. For each task, we report the mean results over five random runs.

\subsection{Cold-start Experiments (RQ1)}\label{sec:cold}
This section presents experimental results and in-depth discussions of PTUPCDR on cold-start scenarios. Following the existing bridge-based methods~\cite{man2017cross,zhu2018deep,fu2019deeply,kang2019semi,cheng2020catn,zhu2021transfer}, we evaluate the performance of PTUPCDR on cold-start scenarios. We demonstrate the effectiveness of PTUPCDR on 3 CDR tasks under different values of $\beta$. As the experimental results are shown in Table~\ref{tab:result}, the best performance is shown in boldface, $*$ indicates $0.05$ level paired t-test of PTUPCDR vs. the best baseline, $Improve$ denotes relative improvement over the best baseline. From the experimental results, we have several findings: (1) TGT is a single-domain model that only uses data from the target domain, and its performance is unsatisfying. Compared with TGT, all other cross-domain methods could exploit data from the source domain, thus achieving better results. Therefore, utilizing data from an auxiliary domain is an effective way to alleviate data sparsity and improve the recommendation performance in the target domain. (2) CMF uses the auxiliary data by combining the data from different domains into a single domain, while CDR methods are specially designed to bridge the domains. We find that CDR methods can outperform CMF in most tasks. It is because CMF ignores the potential domain shift by regarding the data from both domains as the same. On the contrary, the bridge functions can transform the source embeddings into the target feature space, which effectively alleviates the influence of domain shift. Thus, it is essential to study CDR by using the auxiliary domain more effectively. (3) We find that PTUPCDR could outperform the best baseline significantly in most scenarios, which demonstrates that PTUPCDR is effective for cold-start recommendation.

\subsection{More Practical Scenarios (RQ2)}
In this section, we perform experiments and analysis on the compatibility of PTUPCDR with more complicated base models and its utility in the warm-start stage.

\subsubsection{Generalization Experiments:} Note that the bridge-based CDR methods~\cite{man2017cross,zhu2018deep} focus on the bridge function itself, and works in the literature mainly apply their methods upon MF to conduct experimental evaluations. However, MF is a non-neural model, and it is probably too simple to achieve satisfying performance in large-scale real-world recommendations. Thus, to testify the compatibility of PTUPCDR as well as other bridge-based methods, we apply EMCDR and our PTUPCDR upon two more complicated neural models. In other words, we use other models to replace the MF: GMF~\cite{he2017neural} and YouTube DNN~\cite{covington2016deep}. GMF assigns various weights for different dimensions in the dot-product prediction function, which can be regarded as a generalization of vanilla MF. YouTube DNN is a two-tower model. For GMF, the bridge function directly transforms the user embeddings. For YouTube DNN, the bridge function transforms the output of the user tower. For both GMF and YouTube DNN, we train the model with data from both domains. With $\beta = 0.2$, we conduct the generalization experiments on both non-neural (MF) and neural models (GMF, YouTube DNN). Other experimental settings are consistent with Section~\ref{sec:implementation}. From the results shown in Figure~\ref{fig:generalization}(a)(b)(c), we have several insightful observations: (1) The bridge-based CDR methods can be applied upon various base models. With different base models, both EMCDR and PTUPCDR effectively improve the recommendation performance for cold-start users in the target domain. Since GMF and YouTube DNN are two popular and well-designed models in large-scale real-world recommendations, they achieve better performance than the vanilla MF. (2) The generalized PTUPCDR could achieve satisfying performance. On the one hand, with various base models, the generalized PTUPCDR can constantly achieve the best results. On the other hand, as the cold-start problem is highly challenging, the achieved $MAE$ is good enough to testify the effectiveness of generalized PTUPCDR in real-world scenarios.


\subsubsection{Warm-start Experiments}
The existing bridge-based CDR works in the literature~\cite{man2017cross,zhu2018deep,fu2019deeply,kang2019semi,cheng2020catn} only conduct experiments on the extreme cold-start stage. Actually, bridge-based CDR methods are also highly helpful for the warm-start stage by using the mapped embeddings to initialize cold-start users' embeddings of TGT for further training. In real-world recommendations, such warming-up scenarios~\cite{pan2019warm,zhu2021learning} have great application value.

We conduct experiments on TGT, CMF, EMCDR, and our proposed PTUPCDR. In the warm-start training stage, i.e., warm-up process, CMF, EMCDR, and PTUPCDR can be viewed as pre-trained models for initialization. For CMF, we use the collectively trained embeddings to initialize both user and item embeddings of TGT. For EMCDR and PTUPCDR, we initialize the cold-start users' embeddings with mapped users' embeddings. From the results shown in Figure~\ref{fig:warm}, we have the following observations:

\textbf{Cold-start vs. Warm-start.} More interactions can improve the performance of recommender systems. We find that all models in the warm-start stage can achieve better performance than the cold-start stage, demonstrating that more interactions can help recommendation models understand the users better.

\textbf{Utility.} In the warm-start stage, with pre-trained embeddings as the initial embeddings, CMF, EMCDR, PTUPCDR can achieve better performance than TGT, which uses randomly initialized embeddings. PTUPCDR and EMCDR outperform CMF, demonstrating that embeddings pre-trained by bridge-based CDR methods could better exploit the source domain. Such utility of CDR methods on real-world warm-start scenarios is of great practical value.

\textbf{Performance.} In the warm-start stage, our PTUPCDR can still achieve the best results with various test ratio of $\beta$. Therefore, our PTUPCDR is useful and effective in both the cold-start stage and the warm-start stage. 

\subsection{Explanation of the Improvement (RQ3)}

In this section, we conduct extensive experiments and present insightful discussions concerning three modules of PTUPCDR to explain the improvement brought by PTUPCDR and answer $RQ3$. 


\begin{figure}[t]
\centering
\begin{minipage}[b]{1\linewidth}
\centering
\includegraphics[width=0.9\linewidth]{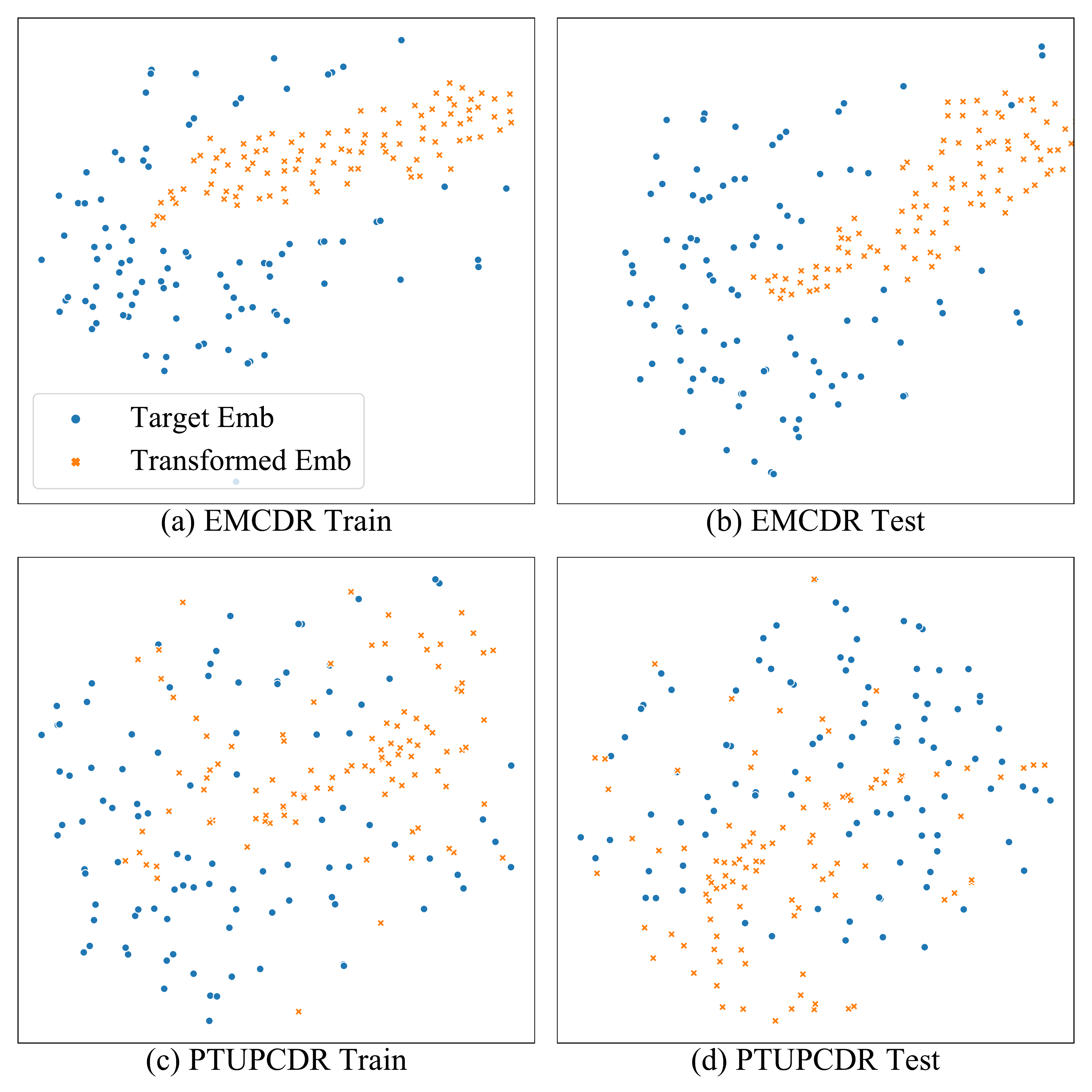}
\end{minipage}
\caption{t-SNE visualization of randomly sampled user embeddings in target-domain feature space and transformed user embeddings. (a) and (b), (c) and (d) denotes the visualization results of EMCDR and PTUPCDR, respectively.}\label{fig:tsne}
\end{figure}

\textbf{Latent Factor Visualization.} We analyze embeddings on the target-domain feature space to further investigate the reason why PTUPCDR outperforms EMCDR and to show the capacity of the Meta Network to generate personalized bridge functions.

We employ the default setting of the t-SNE~\cite{donahue2014decaf} in Scikit-learn to visualize the user embeddings learned by EMCDR and PTUPCDR on Task3 with $\beta = 0.2$. Figure~\ref{fig:tsne} (a) and (b) denote the embeddings of training and test users by EMCDR, while the visualized embeddings in Figure~\ref{fig:tsne} (c) and (d) are learned by PTUPCDR. The blue points denote the target embeddings taken from the target model learned with both training and test users and are regarded as ground truths, while the orange points represent the transformed embeddings. For clarity, we randomly sample 100 training and test users respectively to plot. Note that PTUPCDR and EMCDR share the source and target models, and the only difference is whether the bridge function is personalized by our PTUPCDR or learned by EMCDR.

Ideally, the distributions of the transformed embeddings are the same as the target embeddings. From Figure~\ref{fig:tsne} (a) and (b), we observe that the target embeddings (ground truths) are scattered, while the embeddings transformed by EMCDR are very concentrated. The main reason would be that the single bridge function is hard to capture the complex relationships between users' preferences in the source and target domains. As shown in Figure~\ref{fig:tsne}(c) and (d), PTUPCDR achieves better results. For one thing, the transformed embeddings by PTUPCDR are scattered across the target-domain feature space instead of being clustered as EMCDR, demonstrating the personalization capacity of the Meta Network and the PTUPCDR. More importantly, the distribution of embeddings transformed by PTUPCDR could better fit the target embeddings distributions, which could be the fundamental reason why PTUPCDR could achieve better overall performance. 

\begin{figure}[t]
\centering
\begin{minipage}[b]{1\linewidth}
\centering
\includegraphics[width=1\linewidth]{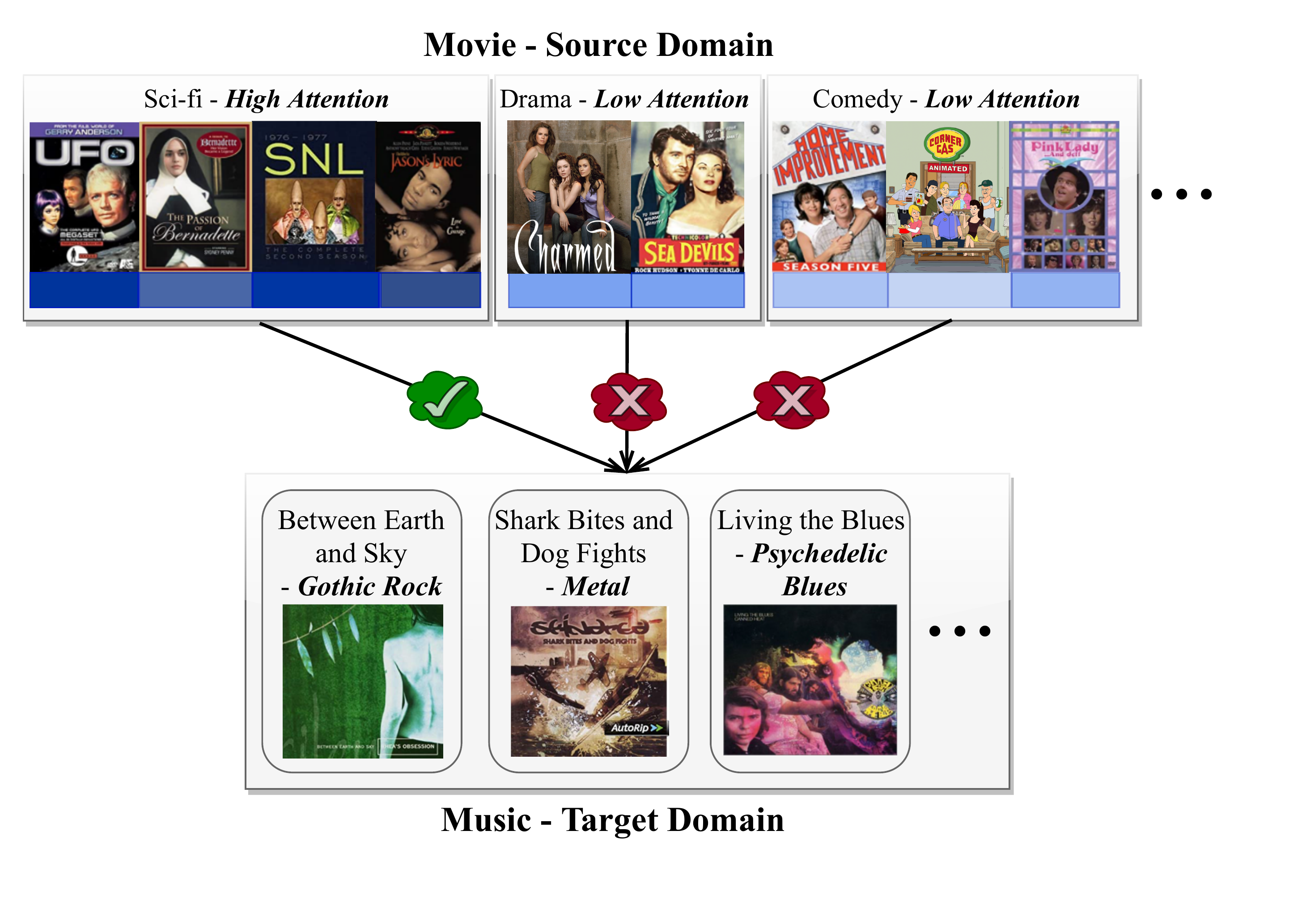}
\end{minipage}\vspace{-0.5cm}
\caption{The color block below each movie represents the attention score. High attention items dominate the recommendation while others have little influence on results.}\label{fig:case_study}
\end{figure}

\textbf{Case Study.}
We present a case study to discuss the necessity and effectiveness of the attention-based Characteristic Encoder. As shown in Figure~\ref{fig:case_study}, our goal is to recommend CDs to a user who has not purchased any CD before, with the help of interacted movies of that user. Note that the darker color block below a movie represents a higher degree of predicted attention over that movie, and the shown 3 CDs are successfully recommended to this user. 

Those successfully recommended hard rock records are somewhat related to interacted science fiction films because they are both exciting. However, hard rock records are almost irrelevant to dramas and comedies. Thus, it is evident that the importance of different interacted items in the source domain should be modeled appropriately. Thus, it is necessary to adopt the attention mechanism to evaluate the items' different contributions to knowledge transfer automatically. At the same time, although the consumed CDs are related to only part of historical interacted movies, the proposed model still could provide relatively accurate recommendations regardless of the influence of noise from dramas and comedies, which demonstrates the effectiveness of the attention-based Characteristic Encoder. To summarize, the attention-based Characteristic Encoder could capture transferable individual characteristics, while existing bridge-based CDR methods ignore this point.
\section{Conclusion}
In this paper, we studied cross-domain recommendation (CDR) which aims to transfer user preferences from an auxiliary domain to the target domain. Many existing CDR methods learn a common preference bridge to transfer preferences. However, a single bridge function shared by all users is hard to capture various relationships between user preferences in source and target domains. Thus, we proposed a novel framework named Personalized Transfer of User Preferences for CDR (PTUPCDR). Specifically, a meta network fed with users' characteristic embeddings is learned to generate personalized bridge functions to achieve personalized transfer of user preferences.  We conducted extensive experiments on real-world datasets to evaluate the proposed PTUPCDR, and the results validate the effectiveness of PTUPCDR on both cold-start and warm-start stages.
\section{Acknowledgments}
\label{sec:ack}
The research work is supported by the National Key Research and Development Program of China under Grant No. 2021ZD0113602,  the National Natural Science Foundation of China under Grant No. U183620661773361, 62176014, U1836206, U1811461.

\bibliographystyle{ACM-Reference-Format}
\balance
\bibliography{main.bbl}

\end{document}